\documentclass[sigconf,nonacm]{acmart}

\AtBeginDocument{%
  \providecommand\BibTeX{{%
    \normalfont B\kern-0.5em{\scshape i\kern-0.25em b}\kern-0.8em\TeX}}}

\usepackage{subcaption}
\usepackage{tikz}
\usepackage{censor}
\usepackage{verbatim}
\usepackage{enumerate}
\usepackage{booktabs}
\usepackage{graphicx}
\usepackage{multirow}
\usepackage{tikz}
\usepackage[htt]{hyphenat}
\newcommand*\circled[1]{\tikz[baseline=(char.base)]{
            \node[white,shape=circle,fill=cyan,draw,inner sep=0.8pt,minimum size=.4cm] (char) {#1};}}
\newcommand{\ignore}[1]{}
\newcommand{\systemName}{Sesame}
\renewcommand{\textcolor}[2]{#2}

\copyrightyear{2021} 
\acmYear{2021} 
\setcopyright{acmlicensed}\acmConference[MAISP'21]{1st Workshop on Security and Privacy for Mobile AI }{June 24, 2021}{Virtual, WI, USA}
\acmBooktitle{1st Workshop on Security and Privacy for Mobile AI (MAISP'21), June 24, 2021, Virtual, WI, USA}
\acmPrice{15.00}
\acmDOI{10.1145/3469261.3469405}
\acmISBN{978-1-4503-8601-2/21/06}

\acmJournal{TOG}
\acmVolume{37}
\acmNumber{4}
\acmArticle{111}
\acmMonth{8}

\begin{document}
\makeatletter
\def\@copyrightspace{\relax}
\makeatother

\title{Open, Sesame!
Introducing Access Control to Voice Services}


\author{Dominika Woszczyk}
\email{d.woszczyk19@imperial.ac.uk}
\affiliation{%
  \institution{Imperial College London}
  \city{London}
  \country{UK}
}

\author{Alvin Lee}
\email{alvin.lee17@imperial.ac.uk}
\affiliation{%
  \institution{Imperial College London}
  \city{London}
  \country{UK}
}

\author{Soteris Demetriou}
\email{s.demetriou@imperial.ac.uk}
\affiliation{%
  \institution{Imperial College London}
  \city{London}
  \country{UK}
}

\renewcommand{\shortauthors}{Woszczyk, et al.}

\begin{abstract}
Personal voice assistants (VAs) are shown to be vulnerable against record--and--replay, and other acoustic attacks which allow an adversary to gain unauthorized control of connected devices within a smart home. Existing defenses either lack detection and management capabilities or are too coarse-grained to enable flexible policies on par with other computing interfaces. In this work, we present \systemName{}, a lightweight framework for edge devices which is the first to enable fine-grained access control of smart-home voice commands. \systemName{}, combines three components: Automatic Speech Recognition, Natural Language Understanding (NLU) and a Policy module.  We implemented \systemName{} on Android devices and demonstrate that our system can enforce security policies for both Alexa and Google Home in real-time (362ms end-to-end inference time),  with a lightweight (<25MB) NLU model which exhibits minimal accuracy loss compared to its non-compact equivalent.
\end{abstract}




\maketitle

\section{Introduction}
\label{sec:intro}

Voice assistants (VAs) such as Alexa or Google Home are gaining popularity as the main means of interaction with various smart-home devices. In December 2019, there were nearly 160 million smart speakers in U.S. homes spread across over 60 million households~\cite{survey}. The introduction of VAs in households comes with serious security challenges as a numerous voice-controlled smart-home devices perform safety and privacy-related operations such as controlling door locks, cameras and home alarms~\cite{shezan2020read}.  Unfortunately, previous work showed the feasibility of acoustic attacks against VAs such as record--and--replay attacks and attacks broadcasting inaudible or hidden voice commands that the VA device captures and executes~\cite{replay,dolphin,chen2020devil,commander,hiddencom}.
To address such issues, researchers proposed acoustic approaches on the VA device such as audio squeezing and audio turbulence~\cite{commander}. Such approaches can raise the bar for the adversary. However, there is still a lack of measures able to detect that the system is under attack which can be very useful for forensics and explainability. More importantly, smart-homes are complex systems with multiple devices of varying capabilities which can have very different security and privacy repercussions. Traditionally, such systems are evolved to be equipped with access control which enables both enforcement and management of flexible security policies. For example, appified software platforms such as Android apps market have introduced the use of permissions to communicate to the users what features each app has access to and to limit the number of privileges given to apps. However, to date there is no such mechanism for voice interfaces and unlike mobile apps, the function or operation being called is not accessible to the user as it is locked behind commercial models. Recently, Shezan et al.~\cite{shezan2020read} proposed a tool for identifying the sensitivity of a voice command however this is still too coarse-grained to enable flexible security policies, as it can only classify voice commands as either action and information-retrieving.


In this work, we propose \systemName{}, a new lightweight access control framework for voice interfaces that enables fine-grained security policy management for voice-controlled smart-home devices. \systemName{} consists of an automatic speech recognition (ASR) model that transcribes the commands which are then analyzed by a natural language understanding (NLU) module which can identify user intent not only at the device-level but also at the granularity of device functionalities. \systemName{} then consults with a policy module for an access decision. Based on a user-defined policy, \systemName{} either allows a non-sensitive command \textcolor{blue}{(e.g. ``Turn on the lights'' or ``Play music'')} to go through or trigger a 2FA authentication step for sensitive commands \textcolor{blue}{(e.g. ``Open the front door'' or ``Arm my home'')}.
 
We develop and evaluate a proof of concept implementation of \systemName{} which is compatible with both the Alexa and Google Home ecosystems, the two market leaders of VAs (53\% of shares~\cite{ReportDe35:online}).\textcolor{blue}{We choose Deepspeech as the automatic speech recognition (ASR) module, performing transcriptions locally and we select state-of-the-art transformer-based language models, Bert and its lightweight form MobileBert, to build our NLU component. We train our models for interfaces and device types classification on datasets of utterances that we collected from Alexa's and Google Home's respective app stores and manually annotated. Finally, we develop a simple access control policy that dictates an \textit{allow} or \textit{block} decision according to the triggered functionality.} We found that our \textcolor{blue}{NLU} system is highly accurate (up to 87.8\% and 84.09\% accuracy on average for Alexa interfaces and capabilities respectively and 98.84\%, 99.57\% for Google Home's traits and devices). \systemName{} can also run in real-time on edge-like devices with 362ms average inference time and negligible memory (138mb) and CPU overhead (25\%).

\ignore{
We implement Deepspeech as the ASR module, performing transcriptions locally and we select state-of-the-art transformer-based language models to build our NLU component. To train our models, we collect datasets of utterances from Alexa's and Google Home's respective app stores which we manually annotate for interfaces and devices types, relying on the developers documentation for each platform. Finally, we develop a simple access control policy that dictates an \textit{allow} or \textit{block} decision according to the triggered functionality. We implement each component on Android devices (Android covers 72\% of the market~\cite{}), and evaluate their efficiency and effectiveness as well as our end-to-end system's performance. We show that our framework performs well on multi-class classification for both ecosystems, even for the smallest models and with very small latency. Our results support \systemName{} as a promising solution for enabling fine-grained access control policies for voice interfaces at the edge.

\textcolor{blue}{
We summarize the contributions of the paper as follows:
\begin{enumerate}
\item We provide fine-grained access control for voice interfaces.
\item The proposed system is lightweight and efficient.
\item We demonstrate a real-world implementation and evaluation with two popular VA, Alexa and Google Home.
\end{enumerate}
}
}

\section{Background}
\label{sec:bgrd}

\subsection{Voice Activated Services}
Users can ask assistants to do voice search, inquiry about the weather, commute time,  broadcast the news and launch party games. They can also ask their voice assistant to control their smart devices within a home environment. VA ecosystems such as Alexa and Google Home provide a collection of ``apps'' for their platform similar to mobile apps. Alexa calls those apps ``skills'' while Google calls them ``actions''. 
 
The VA is passively listening until it hears a wake word (e.g. ``Alexa'' or ``Hey Google''). When detected, the device starts listening actively for a command and streams it to the cloud-hosted Automatic Speech Recognition (ASR) to transcribe it. It then extracts the corresponding intent to call the right interface. For apps that do not control any device, the VA will identify the app name and invoke the corresponding skill/action. 
For smart home devices, both Alexa and Google Home define interfaces that implements different functionalities. Alexa's  documentation divides the functionalities among capabilities which are grouped under higher level interfaces. For instance, the \texttt{Smart Home} interface has capabilities such as \texttt{BrightnessControler} to dim lights or \texttt{TemperatureControler} for the thermostat. 
 Similarly, Google Home defines low-level interfaces named \textit{traits}. Additionally, it also lists different device types. For each trait, the documentation provide a suggestion for corresponding devices that can implement the functionality. For example, the trait \texttt{OpenClose} can be implemented by any device that can open and close such as doors. 

\subsection{Attacks}

Record and replay attacks~\cite{replay} can target the ASR system of voice services in an attempt to trick the speech recognition and authentication algorithms. The underlying technologies can also be subject to acoustic command injections. Cisse et al.~\cite{cisse2017houdini} and Carlini et al.~\cite{hiddencom} were successful in creating inaudible and incomprehensible commands emitted through a compromised speaker in the vicinity of a victim ASR models. Commands were also hidden in songs assuming either white-box~\cite{commander} or black-box~\cite{chen2020devil} access to the ASR model. Finally, voice services are also subject to hardware vulnerabilities. Attackers can exploit the microphone non-linearity distorting sound and pass inaudible commands through an amplifier~\cite{dolphin} but it also has been shown to be possible by using lasers~\cite{sugawara2020light} succeeding in injecting commands from distances of up to 110m and across two buildings.

\section{Threat Model}
\label{sec:threath}
In this work we assume an adversary ($\mathcal{A}$) with full knowledge of the ASR model of a target voice service (white-box). We also assume $\mathcal{A}$ has access to user recordings of voice commands, either through a malicious or compromised mobile app or smart-home device installed on the user's smartphone or smart-home. $\mathcal{A}$ can use these recordings to mount successful record-and-replay attacks on the target ASR model or other acoustic command injection attacks~\cite{cisse2017houdini, hiddencom, commander, chen2020devil, dolphin, sugawara2020light}, to gain unauthorized access to sensitive functionality of smart-home devices such as a smart lock or to probe the indoor activity.


\section{Framework Design}

\begin{figure*}
  \centering
  \includegraphics[width=13cm]{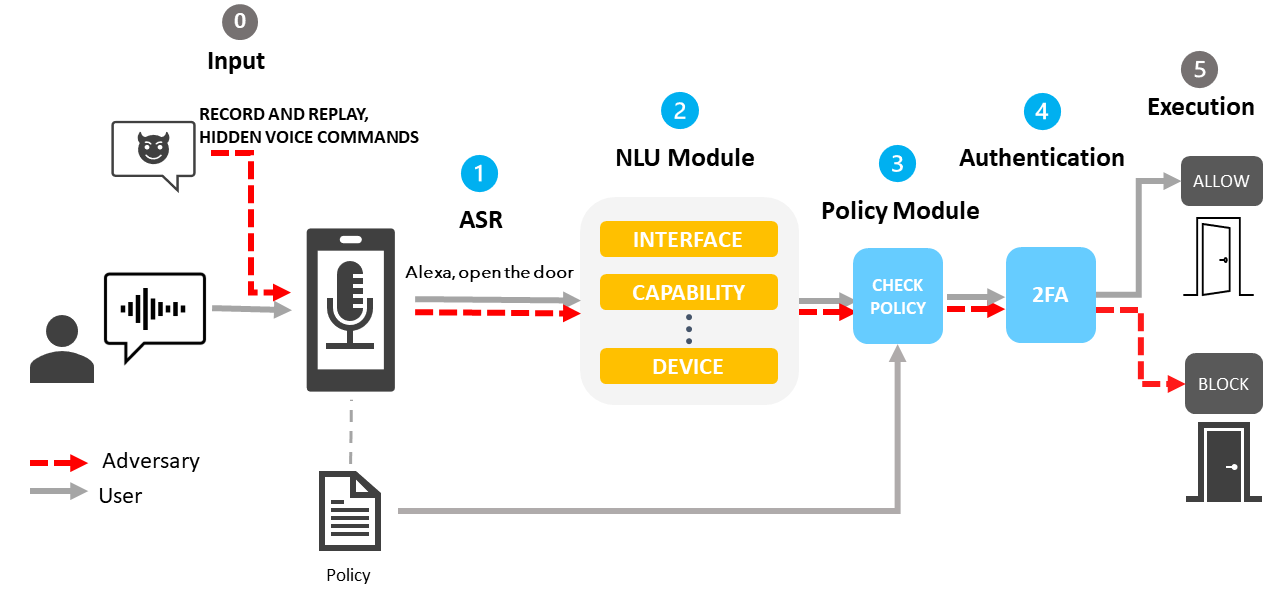}
  \caption{Overview of Sesame, the proposed framework. A spoken command is transcribed by the ASR and classified by the NLU module as a target smart home functionality/device. The Policy module then determines whether the command is to be executed or not and invokes the 2FA and execution steps accordingly.}
  \label{fig:framework}
\end{figure*}

\textcolor{blue}{
In this section, we present the different components of Sesame, as shown on Figure~\ref{fig:framework}. \systemName{} transcribes the command with its ASR module~\circled{1} and the NLU module~\circled{2} detects the intent. The Policy module~\circled{3} then makes the allow or block decision  which invokes the 2FA step~\circled{4} and the access decision is enforced. We describe the modules in the following sections.}
 
\subsection{Automatic Speech Recognition (ASR)}
The first element of the framework is the ASR module  \circled{1} which transcribes the spoken utterances into text. Alexa and Google Home uses their respective cloud-base services (Google's Cloud Speech-to-Text and Amazon Transcribe) before extracting the semantic content. However, this leads to possible additional privacy and security leaks as the user's recordings are sent to the clouds for transcription. Instead, we can perform offline transcriptions with local models and forward the functionality to execute directly to the VA. The ASR model is saved on the device that receives the commands and transcribes incoming utterances locally.

\subsection{Natural Language Understanding (NLU) Module}

Once the utterance is transcribed, the NLU module \circled{2} processes the resulting text and extracts the intended action. In the context of smart homes and voice assistants, intents are "traits", "devices" and "capabilities" that the assistant tries to identify from the transcribed command so it can perform the correct action. Similarly, the NLU component in this framework aims to detect the intent of each command and trigger the appropriate access control rule. 
Intent detection can be performed in a rule--based manner for structured text or with machine learning models for more complex natural language modeling. Most recent language models are based on Transformer models. These are able to model more complex structures and contrary to the rule--based method, there is no need for manual crafting of features.


\subsection{Policy Module}

The Policy module \circled{3} manages access to the different smart home functionalities. To examine the effectiveness of our framework against the attacks considered in Section~\ref{sec:threath}, let us consider a simple system with only two functionalities: un/locking the door and turning on/off the lights. The NLU module is then trained to detect intents corresponding to those tasks from transcribed utterances. We also assume the Policy modules assigns a binary sensitivity class to each intent, "non-sensitive/sensitive", and makes a binary decision, ``allow/block''.
Alice, the user, defines \textit{Lock} as sensitive in their \systemName{} policy and keeps \textit{Light controller} as non-sensitive. Let's consider the following scenarios.

\vspace{5pt}\noindent\textbf{Authorized access to a sensitive activity}. Alice says the command ``Open the door'' (as shown with the continuous arrow on Figure~\ref{fig:framework}). The NLU detects the \textit{Lock} intent and because this action is marked as sensitive by our policy module, the system triggers a second-factor authentication (2FA). Alice authenticate successfully and the smart door opens. 

\vspace{5pt}\noindent\textbf{Unauthorized access to a sensitive activity}. The adversary injects a malicious signal (shown with a dashed arrow on Figure~\ref{fig:framework}) which the ASR transcribes as ``Open the door''. As before, the \textit{Lock} intent is detected and the policy marks it as sensitive. However, when the systems demands confirmation the authentication fails and the command is ignored.


\vspace{5pt}\noindent\textbf{Authorized access to a non-sensitive activity}. 
Alice asks her VA to ``turn on the lamp''. The NLU detects the correct intent, \textit{Lights}, and the policy module marks it as non-sensitive. There is no further action required and the lights turn on. 

\vspace{5pt}\noindent\textbf{Unauthorized access to a non-sensitive activity}. 
Similarly, when an adversary wants to trigger the command ``Turn on the lights'', the framework would identify it as a non-sensitive action and fulfill the command.

The defined policy allows the system to have better usability and less authentication steps. One might change the policy such that all actions required confirmation. However, to maintain a certain usability, the authentication would preferably be performed in a transparent way as mentioned the following Section ~\ref{sec:authentication}. 
Our scenarios also assumes a single user environment. One could introduce a more complex policy module such that it manages the different users within a home and their sensitivity levels preferences.

\subsection{Authentication}
\label{sec:authentication}
Finally, given the output of the policy, the authentication module \circled{4} attempts to identify the user. Currently, Alexa and Google Home provide two-factor authentication (2FA) on potentially sensitive functionalities, but do not enforce it. Alexa provides the option to developers to trigger a spoken PIN while Google can issue a voice challenge. However, a PIN code has to be memorized, can be shared, stolen or overheard. The alternative, voice challenges, is an improvement because the user does not have to remember any specific code. Nevertheless, voice has been shown to be a vulnerable channel for authentication as it can be spoofed~\cite{shamsabadi2020foolhd}.

In this work, we assume that \systemName{} triggers a 2FA step for sensitive functionalities. Instead of using voice authentication, we propose a security first solution and ask the user to confirm with a dialogue box within our framework's app. In future work we would investigate transparent authentication methods that achieve a better usability while preserving security.





\section{Intent Classification}
\label{sec:intent_class}

Smart home utterances often follow a similar structure. The ``wake word'' triggers the assistant and the following words dictate the target app name followed by the intended action and the target device. However, in everyday interactions words can be substituted by synonyms and paraphrases and our NLU model should be able to perform well in those situations. 
In recent years, Transformers and attention-based models have been successfully applied to natural language processing as they are able to model complex syntactic properties. Hence, to model the language in the utterances, we implement two models based on the Transformer architecture. Our baseline model, BERT~\cite{devlin2018bert} has been widely used for state-of-the-art language modeling. However, its base form has a large set of parameters (110M) which impacts both the training and inference time and makes it harder to deploy it on devices with limited resources. Its compact form, MobileBERT~\cite{sun2020mobilebert}, trained using knowledge distillation on a BERT-Large teacher model (320M parameters), reduces the number of parameters to 25M. Additionally, to improve the efficiency of the models and reduce their size, we perform post-training quantization, by converting weights from floating points to floating points of lesser precision. We compare the performance of all four models on the classification tasks of Alexa's interfaces, Alexa's capabilities, Google Home's traits and Google Home's devices. 

\subsection{Dataset Collection}
\label{sec:dataset}


\vspace{5pt}\noindent\textbf{Collecting skills \& actions}.
We built a web crawler to extract skills and their metadata (Name; Store Category; Description; Example  Utterances; Invocation Name) and store the information in a json file. We collected 11,000 skills from the US Alexa store and 5,412 actions from the Google Home store. We remove duplicates and clean the example utterances from specials characters.

\vspace{5pt}\noindent\textbf{Labeling utterances}.
The code of skills is stored in the cloud and it is not openly available. Consequently, we do not have the ground truth about the capabilities triggered by each utterance. To address this, we manually labeled the utterances we collected, using information available on Alexa's~\cite{alexadoc} and Google's Smart Home API documentation~\cite{SmartHom30:online} such as descriptions and example utterances.

\subsection{Datasets}

We collect 1,674 smart home utterances for Alexa and 16,740 for Google Home and add randomly sampled utterances from non-smart home skills and actions, which we group under the \textit{Custom} class. In total our datasets consists of 1,877 Alexa utterances and 25,271 utterances for Google Home.

At the moment of the labeling, Alexa was documenting 10 interfaces and a total of 52 capabilities. We ignored the system calls capabilities and focus on device functionalities. We consider in total 9 interfaces and 33 capabilities. For Google Home, we consider 39 device types and 28 traits.   

\section{Implementation}
\label{sec:implementation}

\vspace{5pt}\noindent\textbf{ASR}. In order to test our work in an end-to-end manner, we introduce an automatic speech recognition system to our implementation. We aim towards a system fully operational on edge, therefore we choose to use the Deepspeech ASR model to transcribe commands. We include Deepspeech (v0.9.3) into our Android app using a pre-trained \textit{tflite} model and interface by Mozilla\footnote{\url{https://github.com/mozilla/DeepSpeech}}, which provides an open-source implementation of a variation of Baidu’s first DeepSpeech paper~\cite{hannun2014deep}.

\vspace{5pt}\noindent\textbf{NLU Inference on Android Device}. We build models using tensorflow tflite\footnote{\url{https://www.tensorflow.org/lite}} packages that provides us with models optimized for mobile devices. We train the models for 15 epochs with early stopping with a batch size of 8 and learning rate of $2\times10^-5$.

\vspace{5pt}\noindent\textbf{Policy}. We implement a simple policy that outputs an allow or block decision.  We define a mapping function from possible intents or devices classes to a Boolean 0 or 1 for block/allow.

\section{Evaluation}
\label{sec:eval}

In our evaluation, we want to analyze the efficiency of our system by measuring the inference latency and usage costs. We also examine the performance of the NLU module in enabling a fine-grained policy management. We aim to answer the following questions:

\begin{enumerate}
\item \textbf{RQ1 (Model Performance):} How accurate is our system at detecting the different levels of interfaces ?

\item \textbf{RQ2 (System Performance):} What is the average inference time and memory use of our systems and its components?

\end{enumerate}

\subsection{Experimental Setup}

We evaluate our framework on an Oneplus 6 with Android 10, 8 GB RAM and a Snapdragon 845 Qualcomm processor. We record the runtime, CPU and memory usage, and the overall accuracy, precision and recall for the classification. For both Alexa and Google Home we use a 85:15 train/test split. The exact size of the train and test sets respectively are 1,586 and 264 for Alexa; 21,440 and 1,384 for Google Home. Duplicated utterances were kept for training but removed from the test sets.

To evaluate the end-to-end system  we synthesize spoken samples from our test utterances using Google Cloud Text-to-Speech \footnote{\url{https://cloud.google.com/text-to-speech}}. We then play the audios through Dell XPS 15 built-ins speakers, back to the microphone of the Android device and save the resulting transcriptions. We then record the prediction performance of the four models on those transcriptions.

\subsection{Classification}

We evaluate the classifier performance for both Alexa and Google. Table~\ref{tab:clf_perf} presents the accuracies for the NLU models when assuming perfect transcriptions, and the end-to-end classification performance with ASR and Policy module on top of the NLU module. We can see that for all four tasks nearly all models achieves average accuracies above 80\% for the NLU-only system. We can also observe that smaller models size and quantization have a negligible impact on the performance compared to their larger model and achieve similar average accuracies. Although there are less classes for Alexa's interfaces and capabilities, the models perform better on Google Home utterances for devices and traits. This can be explained by the small training and test sets for Alexa compared to Google Home's. In terms of end-to-end performance, we can see the that the ASR has a negative impact on the ability of the models to perform the classification. The performance of all models drops considerably for Alexa's capabilities but the impact is negligible for Google Home tasks. Similarly to NLU-only system, the loss of accuracy for smaller models is not significant.

\begin{table}[]
\centering
\caption{Average accuracy for Alexa (A) and Google (G) classification task for NLU-only (NLU) and end-to-end system (E2E). The number in bold is the best value for each column. }
\label{tab:clf_perf}
\resizebox{\columnwidth}{!}{%
\begin{tabular}{|c|c|c|c|c|c|c|c|c|}
\hline
\multirow{2}{*}{Models}            & \multicolumn{2}{c|}{(A) Interface (\%)} & \multicolumn{2}{c|}{(A) Capability (\%)} & \multicolumn{2}{c|}{(G) Traits (\%)}  & \multicolumn{2}{c|}{(G) Devices (\%)} \\ \cline{2-9}
\multicolumn{1}{|l}{} &
  \multicolumn{1}{|c|}{NLU} &
  \multicolumn{1}{c}{E2E} &
  \multicolumn{1}{|c|}{NLU} &
  \multicolumn{1}{c}{E2E} &
  \multicolumn{1}{|c|}{NLU} &
  \multicolumn{1}{c}{E2E} &
  \multicolumn{1}{|c|}{NLU} &
  \multicolumn{1}{c|}{E2E} \\ 
  \hline
BERT             & \textbf{87.88}  & \textbf{79.54}  & \textbf{84.09}   & \textbf{67.04}  & \textbf{98.84} & \textbf{94.79} & \textbf{99.57} & \textbf{94.58} \\
BERT-quant       & 86.74           & 77.65           & 84.09            & 66.66           & 98.77          & 94.58          & 99.56          & 94.65          \\
MobileBERT       & 86.74           & 75.75           & 79.89            & 65.90           & 98.77          & 94.36          & 99.42          & 94.43          \\
MobileBERT-quant & 86.74           & 75.75           & 80.30            & 66.66           & 98.84          & 94.50          & 99.49          & 94.29  
\\
\hline
\end{tabular}
}
\end{table}

\begin{figure*}
     \centering
     \begin{subfigure}[b]{0.245\textwidth}
         \centering
         \includegraphics[width=\textwidth]{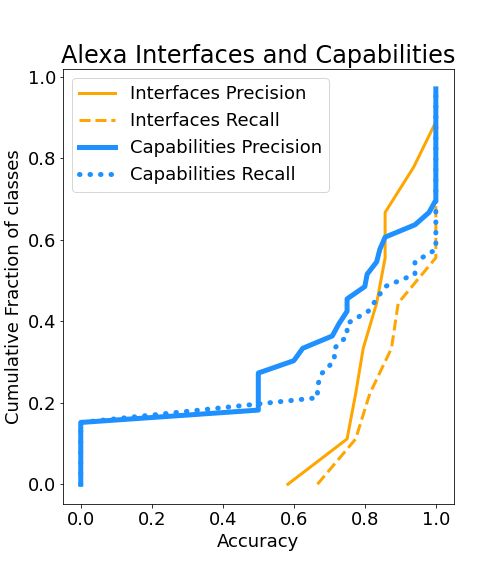}
         \caption{}
         \label{fig:y equals x}
     \end{subfigure}
     \begin{subfigure}[b]{0.245\textwidth}
         \centering
         \includegraphics[width=\textwidth]{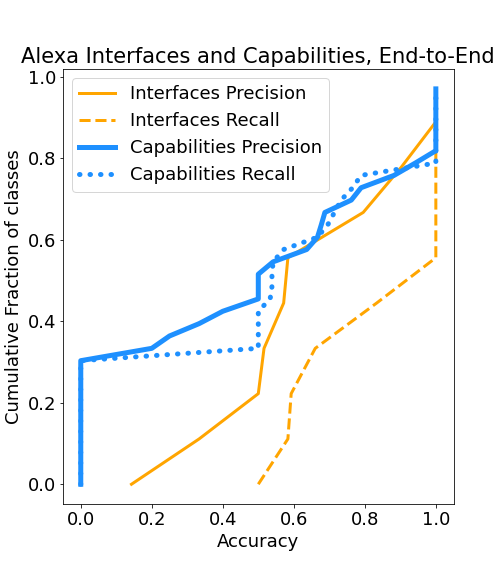}
         \caption{}
         \label{fig:three sin x}
     \end{subfigure}
     \begin{subfigure}[b]{0.245\textwidth}
         \centering
         \includegraphics[width=\textwidth]{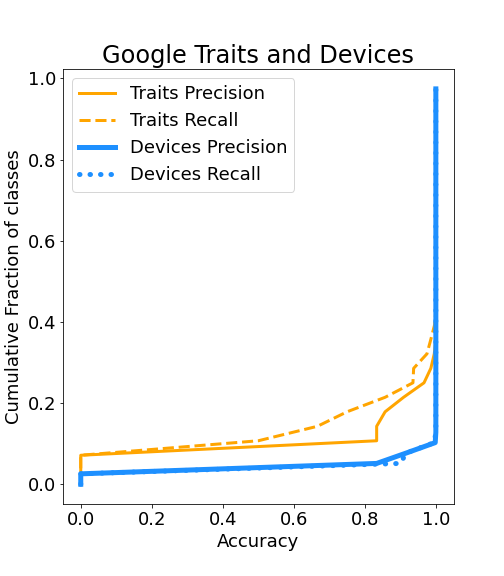}
         \caption{}
         \label{fig:five over x}
     \end{subfigure}
     \begin{subfigure}[b]{0.245\textwidth}
         \centering
         \includegraphics[width=\textwidth]{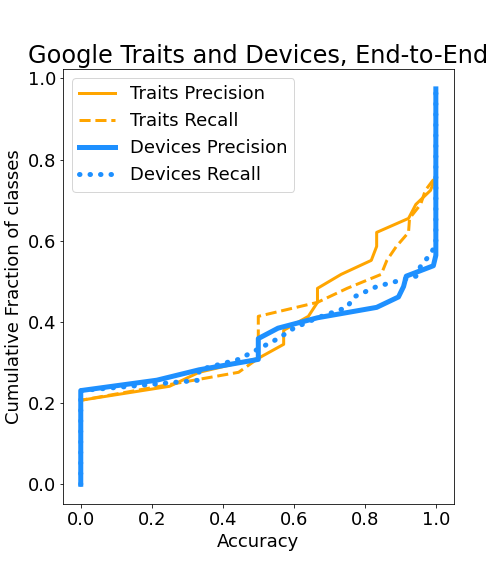}
         \caption{}
         \label{fig:five over x}
     \end{subfigure}
     
     \vspace{-1\baselineskip}
        \caption{Empirical cumulative density functions (ecdf) of precision and recall for the $MobileBERT$-$quant$ model. The 25, 50, 75-percentile for $precision$ are (a) 0.77, 0.83 and 0.86 for Alexa interfaces NLU-only, 0.5, 0.80 and 1 for Alexa capabilities, (b) 0.5, 0.57, 0.80 for interfaces end-to-end (E2E), 0.0, 0.5 and  0.79 for capabilities E2E. For Google traits and devices the percentile are (c) 0.95, 1, 1 and (d) 0.11, 0.67, 0.98 for traits E2E and 0.11, 0.90 and 1 for devices E2E. The percentile for $recall$ are (a) 0.81, 0.89 and 1 for interfaces, 0.67, 0.86 and 1 for capabilities, (b) 0.59, 0.83 and 1 for interfaces E2E, 0.0, 0.54 and 0.77 for capabilities E2E, (c) 0.92, 1, 1 for traits and devices and (d) 0.1, 0.75 and 0.97 for traits E2E and 0.83, 1 and 1 for devices E2E.}
        \label{fig:pr_graphs}
\end{figure*}

On Figure~\ref{fig:pr_graphs}, we compare the ecdf of precision and recall for our smallest model, $MobileBERT$-$quant$, over the tasks' classes to better observe the performance of the NLU module and the impact of the ASR on that performance. Overall, precision and recall for all tasks are similar, although the model has a higher recall. The model performs the worst overall on Alexa's tasks and specifically on capabilities (most classes within 50-100\%). It is even more affected when tested end-to-end (0-79\% with median at 50\%). However, the model performs quite well on the interfaces task (77-80\%) but have a drop in precision on end-to-end (50-80\%). Google traits and devices tasks have very high recall and precision for most classes (92-100\%), with few exception, even with the added transcription error of the ASR (67\% precision and 75\% recall medians for traits and 90\%, 100\% for devices). We investigate classes with low precision and recall and look at the misclassified utterances. Most of the classes with low accuracy are the ones with few train and test samples. The other cases are samples which class is semantically similar to another class that has more labels. 


\begin{table}[]
\centering
\caption{Precision, recall and F1-score for Alexa (A) and Google (G) classification tasks on sensitive interfaces, for NLU-only (NLU) and end-to-end system (E2E).}
\label{tab:clf_perf_sensitive}
\resizebox{\columnwidth}{!}{%
\begin{tabular}{|l|c|c|c|c|c|c|}
\hline
\multirow{2}{*}{Tasks}            & \multicolumn{2}{c|}{Precision (\%)} & \multicolumn{2}{c|}{Recall (\%)} & \multicolumn{2}{c|}{F1-score (\%)} \\ \cline{2-7}
\multicolumn{1}{|l}{} &
  \multicolumn{1}{|c|}{NLU} &
  \multicolumn{1}{c}{E2E} &
  \multicolumn{1}{|c|}{NLU} &
  \multicolumn{1}{c}{E2E} &
  \multicolumn{1}{|c|}{NLU} &
  \multicolumn{1}{c|}{E2E} \\ 
  \hline
(A) Interfaces & 81.39 & 79.54 & 79.54 & 66.03 & 80.46 & 72.16\\
(A) Capabilities  & 75.00 & 70.45 & 82.50 & 70.45 & 78.57 & 70.45\\
(G) Traits  & 98.11 & 71.70 & 98.11 & 90.48 &  98.11 & 80.00 \\
(G) Devices  & 95.83 & 87.50	& 100.00 & 87.50	& 97.87 & 87.50 \\
\hline
\end{tabular}
}
\end{table}
\textcolor{blue}{Finally, we evaluate the precision, recall and overall F1-scores of the $MobileBERT-quant$ classifier on selected ``sensitive'' interfaces. For Alexa, we pick the \textit{Smart Home Security} interface and the corresponding 7 capabilities (controllers for locks, doorbells, camera stream, security system, contact and motion sensor). For Google Home, we select  the traits and devices for which Google's documentation recommends to enable 2FA (\textit{ArmDisarm}, \textit{LockUnlock}, \textit{OpenClose}, and \textit{Lock}, \textit{Gate}, \textit{Garage}, \textit{Door}, \textit{SecuritySystem}, \textit{Network}). The results shown on Table~\ref{tab:clf_perf_sensitive} demonstrate the model is able to classify correctly most of the sensitive utterances for all tasks and also with high precision (80\%, 79\%  F1-score for Alexa interfaces and capabilities, 98\%, 98\% for Google Home’s traits and devices). Although the values for precision and recall drop for E2E, the system recall performance remains high (70\% recall for Alexa capabilities and 90\%, 88\% for Google Home’s traits and devices) with the exception of Alexa interfaces (66\%). The performance of the model could be improved by training it on the sensitivity task directly. However, these results indicates that the model can provide high security on sensitive classes while preserving utility with low number of false positives, which would translate to fewer authentication requests.}  





\subsection{Efficiency}
Table~\ref{tab:system_resources} shows the size, mean inference time, CPU and memory usage over 50 runs of the NLU-only and end-to-end system. We can see an important reduction in runtime and size  between the models and their quantized form (from 1,172 to 244 ms, from 415 to 25 mb)  as well as between the original $BERT$ and $MobileBERT$ models. Although the end-to-end system takes more space overall, the additional inference latency is relatively low, around 110 ms. The CPU and memory usage remains similar for all models except for the $BERT$ model which has higher memory costs.

\begin{table}[]
\centering 
\caption{Mean inference time and memory overhead for NLU-only (NLU) and end-to-end system (E2E). The number in bold is the best value for each column.}
\label{tab:system_resources}
\resizebox{\columnwidth}{!}{%
\begin{tabular}{|c|c|c|c|c|c|c|c|c|}
\hline
\multirow{2}{*}{Models} &
  \multicolumn{2}{c|}{Size (mb)} &
  \multicolumn{2}{c|}{Inference time (ms)} &
  \multicolumn{2}{c|}{CPU (\%)} &
  \multicolumn{2}{c|}{RAM (mb)} \\ \cline{2-9}
\multicolumn{1}{|l|}{} &
  \multicolumn{1}{|c|}{NLU} &
  \multicolumn{1}{c|}{E2E} &
  \multicolumn{1}{|c|}{NLU} &
  \multicolumn{1}{c|}{E2E} &
  \multicolumn{1}{c|}{NLU} &
  \multicolumn{1}{c|}{E2E} &
  \multicolumn{1}{c|}{NLU} &
  \multicolumn{1}{c|}{E2E} \\ 
  \hline 
BERT             & 415  & 1420 & 1172 $\pm$ 11 & 1296 $\pm$ 44 & \textbf{12} & 28 & 440 & 502 \\
BERT-quant       & 105  & 1102 & 717 $\pm$ 3   & 817 $\pm$ 30  & \textbf{12} & 27 & 187 & 241 \\
MobileBERT       & 100  & 1097 & 336 $\pm$ 2   & 446 $\pm$ 30  & \textbf{12} & \textbf{25} & 186 & 201 \\
MobileBERT-quant & \textbf{24.5} & \textbf{1021} & \textbf{244 $\pm$ 10}  & \textbf{362 $\pm$ 22}  & \textbf{12} & \textbf{25} & \textbf{126} & \textbf{138} \\ \hline
\end{tabular}
}
\end{table}

\section{Discussion}
\label{sec:discussion}
Our evaluation shows that our implementation reaches high accuracy over different tasks, with little impact on inference time, memory and CPU usage, answering both \textbf{RQ1} and \textbf{RQ2}. Interestingly, our model performed worse on Alexa's capabilities and interfaces albeit having a smaller number of classes. Combining similar classes would yield better results and less misclassified utterances. A limitation of our approach is that it relies on manually labeled data for specific intents. However, as voice commands datasets get released~\cite{lugosch2019speech}, it will no longer be an issue.

The ASR module can have a large impact on the framework accuracy and total size. However, our experiments presents a baseline performance of the proposed system. Future work would evaluate how cloud-based services compare to a local and open source ASR and their effect on the NLU performance and the security-usability trade-off. Additionally, one might want to skip the ASR step and attempt to identify intents directly from the audio recordings.

Finally, our work provides an example of a possible implementation of \systemName{}. For instance, our policy module is quite simplistic but could be enhanced to handle multiple users and devices efficiently. \systemName{} can be implemented on top of the VA such that there is only one ASR step and it can catch any hidden commands before they are executed by the VA. In future work, we would also investigate more transparent authentication methods.

\section{Related Work}
\label{sec:relwork}
Previous work on access control for IoT enabled smart homes proposed systems that are aware of single and multi-user environments and can make decisions based on context~\cite{demetriou2017hanguard,he2018rethinking,sikder2020kratos}. However, none of these works tackle the challenges of smart home ecosystems controlled  by voice interfaces. \systemName{} adapts to voice-controlled smart homes by introducing the NLU component. 


Shezan et al.~\cite{shezan2020read} perform an empirical study of the sensitivity of the apps available on Alexa Skills store and Google Home Apps store. They build a  natural language processing tool that classifies a given voice command into actions and information retrieval, and define whether the utterance is sensitive or not by checking if it contains any of the selected sensitive keywords. However, this approach offers very limited control in a environment with many smart home devices and requires to manually identify sensitive keywords. 
 
 Spoken language understanding is able to extract intents directly from speech~\cite{lugosch2019speech}.
 However, more experiments need to be done to investigate the latency and performance improvements compared to separate ASR and NLU models. 

\section{Conclusion}
\label{sec:conclusion}
This work presents \systemName{}, an access control framework for voice assistants. \systemName{}'s main components are (1) ASR, (2) Natural Language Processing module, (3) Policy module and (4) Authentication module. We demonstrate the efficiency of the framework on the two most popular voice assistants, Alexa and Google Home, and provide an empirical evaluation on an Android app.   
The NLU module in this paper showed a high intent classification accuracy at different granularity levels, demonstrating the flexibility and precision of the proposed work. Additionally, we show that \systemName{} creates little memory and runtime overhead and that our smallest model perform similarly to their bigger and slower counterparts. 
In future work we plan to explore more sophisticated access control schemes that \systemName{} can support, examine the impact of different model architectures on its performance and the integration of transparent authentication mechanisms.


\bibliographystyle{plainnat}
\setcitestyle{numbers}
\bibliography{lib.bib}

\end{document}